\documentclass[preprintnumbers,amsmath,amssymb,showkeys]{revtex4}
\usepackage{hyperref}
\usepackage[all]{hypcap}
\usepackage{graphicx}
\usepackage{longtable}
\usepackage{color}
\usepackage{url}

\newcommand{\beq}{\begin{eqnarray}}
\newcommand{\eeq}{\end{eqnarray}}
\newcommand{\red}[1]{\textcolor{red}{{#1}}}
\newcommand{\cyan}[1]{\textcolor{blue}{{#1}}}
\newcommand{\darkgreen}[1]{\textcolor{green}{{#1}}}

\newcommand{\Nam}{\mathcal{N}^{m}_{\alpha}}
\newcommand{\bx}{\mathbf{x}}
\newcommand{\by}{\mathbf{y}}
\newcommand{\ba}{\mathbf{a}}
\newcommand{\bv}{\mathbf{v}}

\newcommand{\bff}{\mathbf{f}}
\newcommand{\btheta}{\boldsymbol{\theta}}
\newcommand{\talpha}{\tilde{\alpha}}
\newcommand{\tbv}{\tilde{\mathbf{v}}}
\newcommand{\tpsi}{\tilde{\psi}}

\newcommand{\tW}{\tilde{W}}

\begin{document}

\title{Identifying Hosts of Families of Viruses:\\A Machine Learning Approach}

\author{Anil Raj}
 \affiliation{Department of Applied Physics and Applied Mathematics\\ Columbia University, New York}
 \email{ar2384@columbia.edu}
\author{Michael Dewar}
 \affiliation{Department of Applied Physics and Applied Mathematics\\ Columbia University, New York}
\author{Gustavo Palacios}
 \affiliation{Center for Infection and Immunity\\
            Mailman School of Public Health\\ Columbia University, New York}
\author{Raul Rabadan}
 \affiliation{Department of Biomedical Informatics\\
            Center for Computational Biology and Bioinformatics\\
            Columbia University College  of Physicians and Surgeons, New York}
\author{Chris H. Wiggins}
 \affiliation{Department of Applied Physics and Applied Mathematics\\
                Center for Computational Biology and Bioinformatics\\ Columbia University, New York}

\date{\today}

\begin{abstract}
\section*{Abstract}

Identifying viral pathogens and characterizing their transmission is essential to developing effective public health     measures in response to a pandemic. Phylogenetics, though currently the most popular tool used to characterize the likely host of a virus, can be ambiguous when studying species very distant to known species and when there is very little reliable sequence information available in the early stages of the pandemic. Motivated by an existing framework for representing biological sequence information, we learn sparse, tree-structured models, built from decision rules based on subsequences, to predict viral hosts from protein sequence data using popular discriminative machine learning tools. Furthermore, the predictive motifs robustly selected by the learning algorithm are found to show strong host-specificity and occur in highly conserved regions of the viral proteome.

\end{abstract}


\maketitle

\renewcommand{\thesection}{\arabic{section}}
\section{Introduction}

Emerging pathogens constitute a continuous threat to our society, as it is notoriously difficult to perform a realistic assessment of optimal public health measures when little information on the pathogen is available. Recent outbreaks include the West Nile virus in New York (1999); SARS coronavirus in Hong Kong (2002-2003); LUJO virus in Lusaka (2008); H1N1 influenza pandemic virus in Mexico and the US (2009); and cholera in Haiti (2010). In all these cases, an outbreak of unusual clinical diagnoses triggered a rapid response, and an essential part of this response is the accurate identification and characterization of the pathogen.   

Sequencing is becoming the most common and reliable technique to identify novel organisms. For instance, LUJO was identified as a novel, very distinct virus after the sequence of its genome was compared to other arenaviruses \cite{Briese2009}.  The genome of an organism is a unique fingerprint that reveals many of its properties and past history. For instance, arenaviruses are zoonotic agents usually transmitted from rodents.

Another promising area of research is metagenomics, in which DNA and RNA samples from different environments are sequenced using shotgun approaches. Metagenomics is providing an unbiased understanding of the different species that inhabit a particular niche. Examples include the human microbiome and virome, and the Ocean metagenomics collection \cite{Williamson2008}. It has been estimated that there are more than 600 bacterial species living in the mouth but that only 20\% have been characterized.

Pathogen identification and metagenomic analysis point to an extremely rich diversity of unknown species, where partial genomic sequence is the only information available. The main aim of this work is to develop approaches that can help infer characteristics of an organism from subsequences of its genomic sequence where primary sequence information analysis does not allow us to identify its origin. In particular, our work will focus on predicting the host of a virus from the viral genome.

The most common approach to deduce a likely host of a virus from the viral genome is sequence / phylogenetic similarity (i.e., the most likely host of a particular virus is the one that is infected by related viral species). However, similarity measures based on genomic / protein sequence or protein structure could be misleading when dealing with species very distant to known, annotated species. Other approaches are based on the fact that viruses undergo mutational and evolutionary pressures from the host. For instance, viruses could adapt their codon bias for a more efficient interaction with the host translational machinery or they could be under pressure of deaminating enzymes (e.g. APOBEC3G or HIV infection). All these factors imprint characteristic signatures in the viral genome.  Several techniques have been developed to extract these patterns (e.g., nucleotide and dinucleotide compositional biases, and frequency analysis techniques \cite{Touchon2008}). Although most of these techniques could reveal an underlying biological mechanism, they lack sufficient accuracy to provide reliable assessments. A relatively similar approach to the one presented here is DNA barcoding. Genetic barcoding identifies conserved genomic structures that contain the necessary information for classification. 

Using contemporary machine learning techinques, we present an approach to prediciting the hosts of unseen viruses, based on the amino acid sequences of proteins of viruses whose hosts are well known. Using sequence and host information of known viruses, we learn a multi-class classifier composed of simple sequence-motif based questions (e.g., does the viral sequence contain the motif `DALMWLPD'?) that achieves high prediction accuracies on held-out data. Prediction accuracy of the classifier is measured by the area under the ROC curve, and is compared to a straightforward nearest-neighbour classifier. Importantly (and quite surprisingly), a post-processing study of the highly predictive sequence-motifs selected by the algorithm identifies strongly conserved regions of the viral genome, facilitating biological interpretation.

\section{Methods}     

Our overall aim is to discover aspects of the relationship between a virus and its host. Our approach is to develop a model that is able to predict the host of a virus given its sequence; those features of the sequence that prove most useful are then assumed to have a special biological significance. Hence, an ideal model is one that is parsimonious and easy to interpret, whilst incorporating combinations of biologically relevant features. In addition, the interpretability of the results is improved if we have a simple learning algorithm which can be straightforwardly verified.

Formally, for a given virus family, we learn a function $g:\mathcal{S} \rightarrow\mathcal{H}$, where $\mathcal{S}$ is the space of viral sequences and $\mathcal{H}$ is the space of viral hosts. The space of viral sequences $\mathcal{S}$ is generated by an alphabet $\mathcal{A}$ where, $|\mathcal{A}| = 4$ (genome sequence) or $|\mathcal{A}| = 20$ (primary protein sequence).

Defining a function on a sequence requires representation of the sequence in some feature space. Below, we specify a representation $\phi:\mathcal{S}\rightarrow\mathcal{X}$, where a sequence $s \in \mathcal{S}$ is mapped to a vector of counts of subsequences $x \in \mathcal{X} \subset \mathbb{N}_0^D$. Given this representation, we have the well-posed problem of finding a function ${f}:\mathcal{X} \rightarrow \mathcal{H}$ built from a space of simple binary-valued functions.

\subsection{Collected Data}

The collected data consist of $N$ genome sequences or primary protein sequences, denoted $s_1 \ldots s_N$, of viruses whose host class, denoted $h_1 \ldots h_N$ is known. For example, these could be `plant', `vertebrate' and `invertebrate'. The label for each virus is represented numerically as $\by \in \mathcal{Y}=\{0,1\}^L$ where $[\by]_l=1$ if the index of the host class of the virus is $l$, and where $L$ denotes the number of host classes. Note that this representation allows for a virus to have multiple host classes. Here and below we use boldface variables to indicate vectors and square brackets to denote the selection of a specific element in the vector, e.g., $[\by_n]_l$ is the $l^\mathrm{th}$ element of the $n^\mathrm{th}$ label vector. 

\subsection{Mismatch Feature Space}

A possible feature space representation of a viral sequence is the vector of counts of exact matches of all possible $k$-length subsequences ($k$-mers). However, due to the high mutation rate of viral genomes \cite{Duffy2008,Pybus2009}, a predictive function learned using this simple representation of counts would fail to generalize well to new viruses. Instead, motivated by \cite{Leslie2004}, we count not just the presence of an individual $k$-mer but also the presence of subsequences within $m$ mismatches from that $k$-mer. The mismatch- or $m$-neighborhood of a $k$-mer $\alpha$, denoted $\Nam$, is the set of all $k$-mers with a Hamming distance \cite{Hamming1950} at most $m$ from it, as shown in Table \ref{tab:neighborhood}. Let $\delta_{\Nam}$ denote the indicator function of the $m$-neigbourhood of $\alpha$ such that 
\begin{equation}
	\delta_{\Nam}(\beta) = \left\{ 
	\begin{array}{rl}
		1 & \mathrm{if} ~ \beta \in \Nam \\
		0 & \mathrm{otherwise}. 
	\end{array}
	\right. 
\end{equation}

\begin{table}[!ht]
    \caption{{\bf Mismatch feature space representation.} The mismatch feature space representation of a segment of a protein sequence ...\red{AQGPR}IYDDT\cyan{CQHPS}WWMNFE\darkgreen{YRGSP}...}
    \begin{center}
    \begin{tabular}{ cc | cc | cc }
        \hline
        \multicolumn{2}{c}{$m=0$} & \multicolumn{2}{c}{$m=1$} & \multicolumn{2}{c}{$m=2$} \\
        \hline
        kmer & count & kmer & count & kmer & count \\
        $\vdots$ & $\vdots$ & $\vdots$ & $\vdots$ & $\vdots$ & $\vdots$ \\
        DQGPS & 0 & DQGPS & 0 & \cyan{DQGPS} & \cyan{1} \\
        CQGPS & 0 & \cyan{CQGPS} & \cyan{1} & \cyan{CQGPS} & \cyan{1} \\
        \cyan{CQHPS} & \cyan{1} & \cyan{CQHPS} & \cyan{1} & \cyan{CQHPS} & \cyan{1} \\
        CQIPS & 0 & \cyan{CQIPS} & \cyan{1} & \cyan{CQIPS} & \cyan{1} \\
        DQIPS & 0 & DQIPS & 0 & \cyan{DQIPS} & \cyan{1} \\
        $\vdots$ & $\vdots$ & $\vdots$ & $\vdots$ & $\vdots$ & $\vdots$ \\
        APGPQ & 0 & APGPQ & 0 & \red{APGPQ} & \red{1} \\
        AQGPQ & 0 & \red{AQGPQ} & \red{1} & \red{AQGPQ} & \red{1} \\
        \red{AQGPR} & \red{1} & \red{AQGPR} & \red{1} & \red{AQGPR} & \red{1} \\
        AQGPS & 0 & \red{AQGPS} & \red{1} & \red{AQGPS} & \red{1} \\
        ASGPS & 0 & ASGPS & 0 & \red{ASGPS} & \red{1} \\
        $\vdots$ & $\vdots$ & $\vdots$ & $\vdots$ & $\vdots$ & $\vdots$ \\
        ARGMP & 0 & ARGMP & 0 & \darkgreen{ARGMP} & \darkgreen{1} \\
        ARGSP & 0 & \darkgreen{ARGSP} & \darkgreen{1} & \darkgreen{ARGSP} & \darkgreen{1} \\
        \darkgreen{YRGSP} & \darkgreen{1} & \darkgreen{YRGSP} & \darkgreen{1} & \darkgreen{YRGSP} & \darkgreen{1} \\
        WRGSP & 0 & \darkgreen{WRGSP} & \darkgreen{1} & \darkgreen{WRGSP} & \darkgreen{1} \\
        WRGNP & 0 & WRGNP & 0 & \darkgreen{WRGNP} & \darkgreen{1} \\
        $\vdots$ & $\vdots$ & $\vdots$ & $\vdots$ & $\vdots$ & $\vdots$ \\
    \end{tabular}
    \end{center}
    \label{tab:neighborhood}
\end{table}

We can then define, for any possible $k$-mer $\beta$, the mapping $\phi$ from the sequence $s$ onto a count of the elements in $\beta$'s $m$-neighbourhood as
\begin{equation}
	\phi_{k,m}(s,\beta) = \sum_{\substack{\alpha \in s\\|\alpha|=k}} \delta_{\Nam}(\beta).
\end{equation}
Finally, the $d^\mathrm{th}$ element of the feature vector for a given sequence is then defined elementwise as
\begin{equation}
    [\bx]_d = \phi_{k,m}(s,\beta_d)
\end{equation}
for every possible $k$-mer $\beta_d \in \mathcal{A}^k$, where $d = 1\dots D$ and $D = |\mathcal{A}^k|$. 

Note that when $m=0$, $\phi_{k,0}$ exactly captures the simple count representation described earlier. This biologically realistic relaxation allows us to learn discriminative functions that better capture rapidly mutating and yet functionally conserved regions in the viral genome facilitating generalization to new viruses. 

\subsection{Alternating Decision Trees}

Given this representation of the data, we aim to learn a discriminative function that maps features $\bx$ onto host class labels $\by$, given some training data $\{(\bx_1,\by_1), \dots, (\bx_N,\by_N) \}$. We want the discriminative function to output a measure of ``confidence'' \cite{Schapire1999} in addition to a predicted host class label. To this end, we learn on a class of functions $\bff:\mathcal{X}\rightarrow \mathbb{R}^L $, where the indices of positive elements of $\bff(\bx)$ can be interpreted as the predicted labels to be assigned to $\bx$ and the magnitudes of these elements to be the confidence in the predictions. 

A simple class of such real-valued discriminative functions can be constructed from the linear combination of simple binary-valued functions $\psi: \mathcal{X}\rightarrow \{0,1\}$. The functions $\psi$ can, in general, be a combination of single-feature decision rules or their negations: 
\begin{eqnarray}
	\bff(\bx) & = & \sum_{p=1}^P \mathbf{a}_p \psi_p(\bx) \\
	\psi_p(\bx) & = & \prod_{d \in S_p} \mathbb{I}(x_d \geq \theta_d)
\end{eqnarray}
where $\ba_p \in \mathbb{R}^L$, $P$ is the number of binary-valued functions, $\mathbb{I}(\cdot)$ is 1 if its argument is true, and zero otherwise, $\theta \in \{0,1,\dots,\Theta\}$, where $\Theta = \max_{d,n} [\bx_{n}]_d$, and $S_p$ is a subset of feature indices. This formulation allows functions to be constructed using combinations of simple rules. For example, we could define a function $\psi$ as the following
\begin{equation}
	\psi(\bx) = \mathbb{I}(x_5 \geq 2) \times \lnot\mathbb{I}(x_{11} \geq 1) \times \mathbb{I}(x_1 \geq 4)
\end{equation}
where $\lnot\mathbb{I}(\cdot) = 1 - \mathbb{I}(\cdot)$.

Alternatively, we can view each function $\psi_p$ to be parameterized by a vector of thresholds $\btheta_p \in \{0,1,\dots,\Theta\}^D$, where $[\btheta_p]_d = 0$ indicates $\psi_p$ is not a function of the $d^\mathrm{th}$ feature $[\bx]_d $. In addition, following \cite{Busa2009}, we can decompose the weights $\ba_p = \alpha_p \bv_p$ into a vote vector $\bv \in \{+1,-1\}^L$ and a scalar weight $\alpha \in \mathbb{R}_+$. The discriminative model, then, can be written as
\begin{eqnarray}
	\bff(\bx) & = & \sum_{p=1}^P \alpha_p \bv_p \psi_{\btheta_p}(\bx), \\
	\psi(\bx;\btheta_p) & = & \prod_{d=1}^D \mathbb{I}(x_d \geq [\btheta_p]_d).
\end{eqnarray}

\begin{figure}[!ht]
    \begin{center}
        \includegraphics[height=4in]{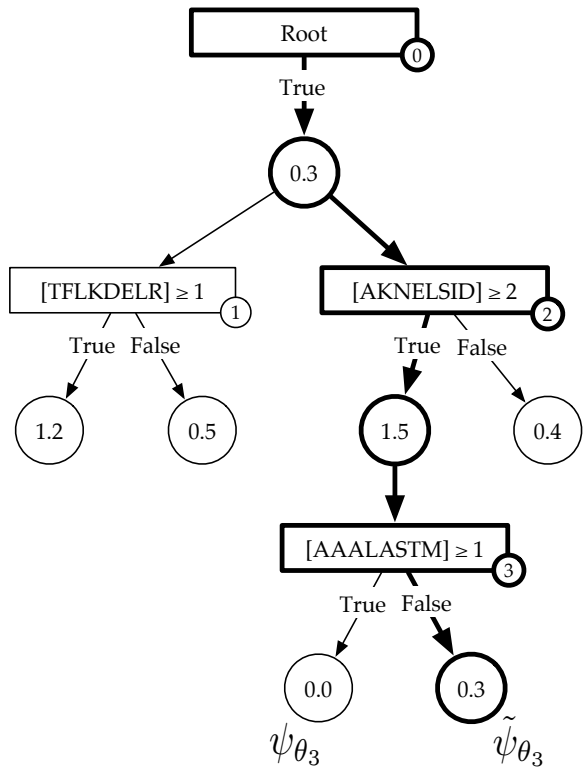}
    \end{center}
    \caption{{\bf Alternating Decision Tree.} An example of an ADT where rectangles are decision nodes, circles are output     nodes and, in each decision node, $[\beta] = \phi_{k,m}(s,\beta)$ is the feature associated with the $k$-mer $\beta$ in        sequence $s$. The output nodes connected to each decision node are associated with a pair of binary-valued functions $(\psi,   \tpsi)$. The binary-valued function corresponding to the highlighted path is given as $\tpsi(\bx;\btheta_3) =                  \mathbb{I}([\mathrm{AKNELSID}] \geq 2) \times \lnot\mathbb{I}([\mathrm{AAALASTM}] \geq 1)$ and the associated $\talpha = 0.3$.}
    \label{fig:adt}
\end{figure}

Every function in this class of models can be concisely represented as an Alternating Decision Tree (ADT) \cite{Freund1999}. Similar to ordinary decision trees, ADTs have two kinds of nodes: decision nodes and output nodes. Every decision node is associated with a single-feature decision rule, the attributes of the node being the relevant feature and corresponding threshold. Each decision node is connected to two output nodes corresponding to the associated decision rule and its negation. Thus, binary-valued functions in the model come in pairs $(\psi, \tpsi)$; each pair is associated with the the pair of output nodes for a given decision node in the tree (see Figure \ref{fig:adt}). Note that $\psi$ and $\tpsi$ share the same threshold vector $\btheta$ and only differ in whether they contain the associated decision rule or its negation. The attributes of the output node pair are the vote vectors $(\bv,\tbv)$ and the scalar weights $(\alpha,\talpha)$ associated with the corresponding functions $(\psi, \tpsi)$.

Each function $\psi$ has a one-to-one correspondence with a path from the root node to its associated output node in the tree; the single-feature decision rules in $\psi$ being the same as those rules associated with decision nodes in the path, with negations applied appropriately. Combinatorial features can, thus, be incorporated into the model by allowing for trees of depth greater than 1. Including a new function $\psi$ in the model is, then, equivalent to either adding a new path of decision and output nodes at the root node in the tree or growing an existing path at one of the existing output nodes. This tree-structured representation of the model will play an important role in specifying how Adaboost, the learning algorithm, greedily searches over an exponentially large space of binary-valued functions. It is important to note that, unlike ordinary decision trees, each example runs down an ADT through every path originating from the root node. 

\subsection{Multi-class Adaboost}

Having specified a representation for the data and the model, we now briefly describe Adaboost, a large-margin supervised learning algorithm which we use to learn an ADT given a data set. Ideally, a supervised learning algorithm learns a discriminative function $\bff^*(\bx)$ that minimizes the number of mistakes on the training data, known as the Hamming loss \cite{Hamming1950}:
\begin{equation}
	\bff^*(\bx) = \arg \min_{\bff} \mathcal{L}_h(\bff) = \sum_{\substack{ 1\le n \le N \\
	1\le l \le L}} \mathbb{I} \left (H([\bff(\bx_n)]_l) \neq [\by_n]_l \right )
\end{equation}
where $H(.)$ denotes the Heaviside function. The Hamming loss, however, is discontinuous and non-convex, making optimization intractable for large-scale problems. 

Adaboost is the unconstrained minimization of the exponential loss, a smooth, convex upper-bound to the Hamming loss, using a coordinate descent algorithm. 
\begin{equation}
	\tilde{\bff}^*(\bx) = \arg \min_{\bff} \mathcal{L}_e(\bff) = \sum_{\substack{ 1\le n \le N \\
	1\le l \le L}} \exp \left (-[\by_n]_l [f_l(\bx_n)]_l \right ). 
\end{equation}
Adaboost learns a discriminative function $\bff(\bx)$ by iteratively selecting the $\psi$ that maximally decreases the exponential loss. Since each $\psi$ is parameterized by a $D$-dimensional vector of thresholds $\btheta$, the space of functions $\psi$ is of size $O((\Theta+1)^D)$, where $\Theta$ is the largest $k$-mer count observed in the data, making an exhaustive search at each iteration intractable for high-dimensional problems. 

To avoid this problem, at each iteration, we only allow the ADT to grow by adding one decision node to one of the existing output nodes. To formalize this, let us define $Z(\btheta) = \{d: [\btheta]_d \neq 0\}$ to be the set of active features corresponding to a function $\psi$. At the $t^\mathrm{th}$ iteration of boosting, the search space of possible threshold vectors is then given as $\{\btheta: \exists \tau<t, Z(\btheta) \supset Z(\btheta_{\tau}), |Z(\btheta)|-|Z(\btheta_{\tau})| = 1\}$. In this case, the search space of thresholds at the $t^\mathrm{th}$ iteration is of size $O(t \Theta D)$ and grows linearly in a greedy fashion at each iteration (see Figure \ref{fig:adt}). Note, however, that this greedy search, enforced to make the algorithm tractable, is not relevant when the class of models are constrained to belong to ADTs of depth 1.

In order to pick the best function $\psi$, we need to compute the decrease in exponential loss admitted by each function in the search space, given the model at the current iteration. Formally, given the model at the $t^\mathrm{th}$ iteration, denoted $\bff^t(\bx)$, the exponential loss upon inclusion of a new decision node, and hence the creation of two new paths $(\psi_{\btheta},\tpsi_{\btheta})$, into the model can be written as
\begin{eqnarray}
	\mathcal{L}_e(\bff^{t+1}) & = & \sum_{n=1}^N \sum_{l=1}^L \exp \left( -[\by_n]_l [\bff^t(\bx_n) 
	+ \alpha \bv \psi_{\btheta}(\bx_n) + \talpha \tbv \tpsi_{\btheta}(\bx_n)]_l \right)  \\
        & = & \sum_{n=1}^N \sum_{l=1}^L w^t_{nl} \exp \left ( -[\by_n]_l [\alpha \bv \psi_{\btheta} (\bx_n) + \talpha \tbv \tpsi_{\btheta}(x_n) ]_l \right ) 
\end{eqnarray}
where $w^t_{nl} = \exp \left ( -[\by_n]_l [\bff^t(\bx_n)]_l \right )$. Here $w_{nl}^t$ is interpreted as the weight on each sample, for each label, at boosting round $t$. If, at boosting round $t-1$, the model disagrees with the true label $l$ for sample $n$, then $w_{nl}^t$ is large. If the model agrees with the label then the weight is small. This ensures that the boosting algorithm chooses a decision rule at round $t$, preferentially discriminating those examples with a large weight, as this will lead to the largest reduction in $\mathcal{L}_e$.

For every possible new decision node that can be introduced to the tree, Adaboost finds the ($\alpha$,$\bv$) pair that minimizes the exponential loss on the training data. These optima can be derived as 
\begin{equation}
	[\bv^*]_l = 
	\begin{cases}
		1 & \text{if } \omega^t_{+,l} \geq \omega^t_{-,l} \\
		-1 & \text{otherwise}
	\end{cases}
    \label{eq:vote}
\end{equation}
\begin{equation}
	\alpha^* = \frac{1}{2} \ln \frac{W^t_+}{W^t_-} 
    \label{eq:weight}
\end{equation}
where for each new path $\psi_n$ associated with each new decision node
\begin{eqnarray}
	\omega^t_{\pm,l} & = & \sum_{n:\psi_n y_{nl}=\pm1} w^t_{nl}\\
	W^t_{\pm} & = & \sum_{n,l:v_l \psi_n y_{nl}=\pm1} w^t_{nl}.
\end{eqnarray}
Corresponding equations for the ($\talpha$,$\tbv$) pair can be written in terms of $\tW^t_{\pm,l}$ and $\tW^t_{\pm}$ obtained by replacing $\psi_n$ with $\tpsi_n$ in the equations above. The minimum loss function for the threshold $\btheta$ is then given as 
\begin{equation}
	\mathcal{L}_e(\bff^{t+1}) = 2 \sqrt{W^t_+ W^t_-} + 2 \sqrt{\tW^t_+ \tW^t_-} + W^t_o
    \label{eq:loss}
\end{equation}
where $W^t_o = \sum_{n,l:\psi_n=\tpsi_n=0} w^t_{nl}$. Based on these model update equations, each iteration of the Adaboost algorithm involves building the set of possible binary-valued functions to search over, selecting the one for which the loss function given by Eq. \ref{eq:loss} and computing the associated $(\alpha,\bv)$ pair using Eq. \ref{eq:vote} and Eq. \ref{eq:weight}.

\section{Results}

\subsection{Data specifications} 
We aim to learn a predictive model to identify hosts of viruses belonging to a specific family; we show results for \emph{Picornaviridae} and \emph{Rhabdoviridae}. \emph{Picornaviridae} is a family of viruses that contain a single stranded, positive sense RNA. The viral genome usually contains about 1-2 Open Reading Frames (ORF), each coding for protein sequences about 2000-3000 amino acids long. \emph{Rhabdoviridae} is a family of negative sense single stranded RNA viruses whose genomes typically code for five different proteins: large protein (L), nucleoprotein (N), phosphoprotein (P), glycoprotein (G), and matrix protein (M). The data consist of 148 viruses in the \emph{Picornaviridae} family and 50 viruses in the \emph{Rhabdoviridae} family. For some choice of $k$ and $m$, we represent each virus as a vector of counts of all possible $k$-mers, up to $m$-mismatches, generated from the amino-acid alphabet. Each virus is also assigned a label depending on its host: vertebrate / invertebrate / plant in the case of \emph{Picornaviridae}, and animal / plant in the case of \emph{Rhabdoviridae} (see Table S1 for the names and label assignments of viruses). Using multiclass Adaboost, we learn an ADT classifier on training data drawn from the set of labeled viruses and test the model on the held-out viruses. 

\subsection{BLAST Classifier accuracy}

Given whole protein sequences, a straightforward classifier is given by a nearest-neighbour approach based on the Basic Local Alignment Search Tool (BLAST) \cite{Altschul1990}. We can use BLAST score (or $P$-value) as a measure of the distances between the unknown virus and a set of viruses with known hosts. The nearest neighbor approach to classification then assigns the host of the closest virus to the unknown virus. Intuitively, as this approach uses the whole protein to perform the classification, we expect the accuracy to be very high. This is indeed the case -- BLAST, along with a $1$-nearest neighbor classifier, successfully classifies all viruses in the \emph{Rhabdoviridae} family, and all but 3 viruses in the \emph{Picornaviridae} family. What is missing from this approach, however, is the ability to ascertain and interpret host relevant motifs.

\subsection{ADT Classifier accuracy} 

The accuracy of the ADT model, at each round of boosting, is evaluated using a multi-class extension of the Area Under the Curve (AUC). Here the `curve' is the Receiver Operating Characteristic (ROC) which traces a measure of the classification accuracy of the ADT for each value of a real-valued discrimination threshold. As this threshold is varied, a virus is considered a true (or false) positive if the prediction of the ADT model for the true class of that protein is greater (or less) than the threshold value. The ROC curve is then traced out in True Positive Rate -- False Positive Rate space by changing the threshold value and the AUC score is defined as the area under this ROC curve. 

\begin{figure}[!ht]
    \begin{center}
        \includegraphics[width=0.7\textwidth]{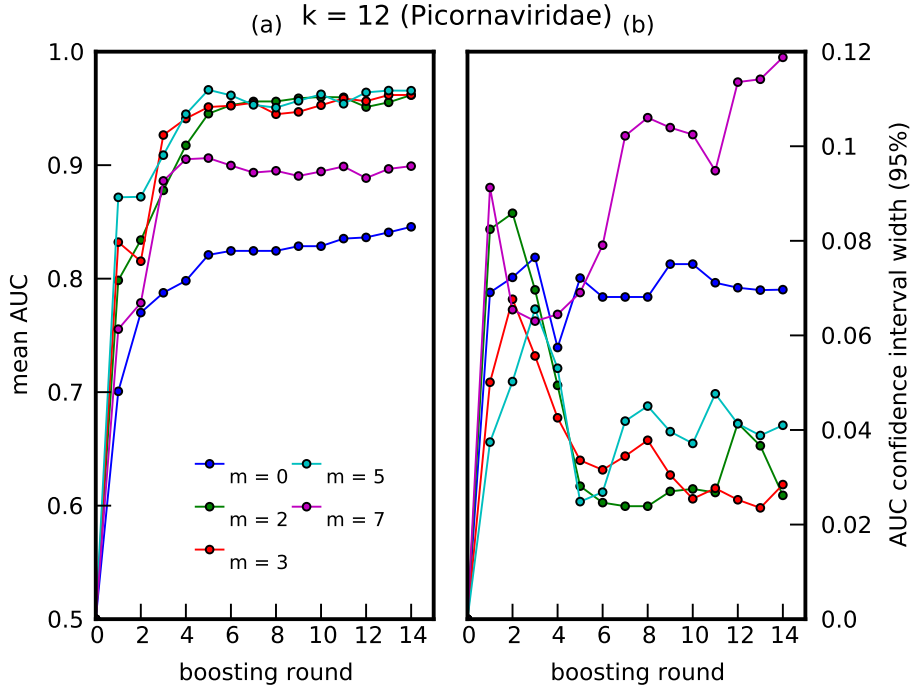}
    \end{center}
    \caption{{\bf Prediction accuracy for \emph{Picornaviridae}.} A plot of (a) mean AUC vs boosting round, and (b) 95\%       confidence interval vs boosting round. The mean and standard deviation were computed over 10-folds of held-out data, for       \emph{Picornaviridae}, where $k = 12$.}
    \label{fig:auc_v_round_picorna}
\end{figure}

\begin{figure}[!ht]
    \begin{center}
        \includegraphics[width=0.7\textwidth]{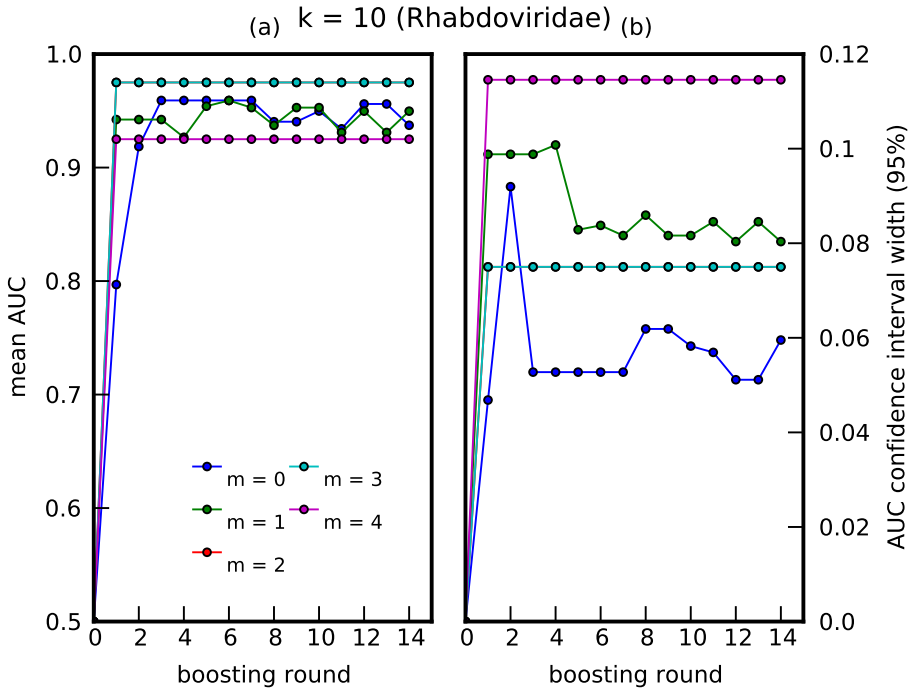}
    \end{center}
    \caption{{\bf Prediction accuracy for \emph{Rhabdoviridae}.} A plot of (a) mean AUC vs boosting round, and (b) 95\%        confidence interval vs boosting round. The mean and standard deviation were computed over 5-folds of held-out data, for        \emph{Rhabdoviridae}, where $k = 10$. The relatively higher uncertainty for this virus family was likely due to very small     sample sizes. Note that the cyan curve lies on top of the red curve.}
    \label{fig:auc_v_round_rhabdo}
\end{figure}

The ADT is trained using 10-fold cross validation, calculating the AUC, at each round of boosting, for each fold using the held-out data. The mean AUC and standard deviation over all folds is plotted against boosting round in Figures \ref{fig:auc_v_round_picorna}, \ref{fig:auc_v_round_rhabdo}. Note that the `smoothing effect' introduced by using the mismatch feature space allows for improved prediction accuracy for $m > 0$. For \emph{Picornaviridae}, the best accuracy is achieved at $m = 5$, for a choice of $k = 12$; this degree of `smoothing' is optimal for the algorithm to capture predictive amino-acid subsequences present, up to a certain mismatch, in rapidly mutating viral protein sequences. For \emph{Rhabdoviridae}, near perfect accuracy is achieved with merely one decision rule, i.e., \emph{Rhabdoviridae} with plant or animal hosts can be distinguished based on the presence or absence of one highly conserved region in the L protein. 

\subsection{Predictive subsequences are conserved within hosts} 

\begin{figure}[!ht]
    \begin{center}
        \includegraphics[width=5in]{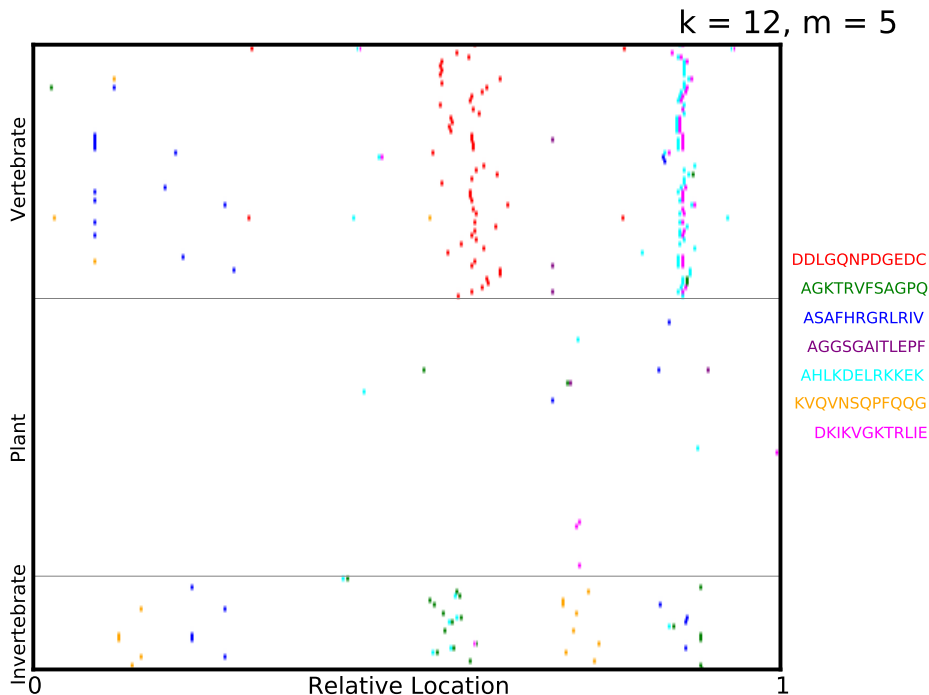}
    \end{center}
    \caption{{\bf Visualizing predictive subsequences.} A visualization of the mismatch neighborhood of the first 7 $k$-mers   selected in an ADT for \emph{Picornaviridae}, where $k = 12, m = 5$. The virus proteomes are grouped vertically by their label with their lengths scaled to $[0,1]$. Regions containing elements of the mismatch neighborhood of each $k$-mer are then        indicated on the virus proteome. Note that the proteomes are not aligned along the selected $k$-mers but merely stacked        vertically with their lengths normalized.}
    \label{fig:predictive_regions}
\end{figure}

\begin{figure}[!ht]
    \begin{center}
        \includegraphics[width=4in]{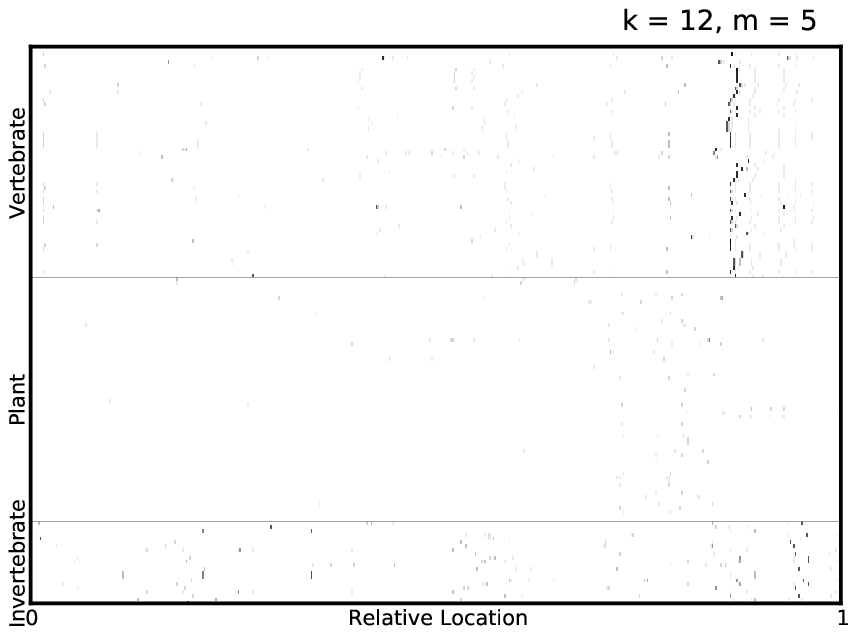}
    \end{center}
    \caption{{\bf Visualizing predictive regions of protein sequences.} A visualization of the mismatch neighborhood of the    first 7 $k$-mers, selected in all ADTs over 10-fold cross validation, for \emph{Picornaviridae}, where $k = 12, m = 5$.        Regions containing elements of the mismatch neighborhood of each selected $k$-mer are indicated on the virus proteome, with    the grayscale intensity on the plot being inversely proportional to the number of cross-validation folds in which some $k$-mer in that region was selected by Adaboost. Thus, darker spots indicate that some $k$-mer in that part of the proteome was        robustly selected by Adaboost. Furthermore, a vertical cluster of dark spots indicate that region, selected by Adaboost to be  predictive, is also strongly conserved among viruses sharing a common host type.}
    \label{fig:robust_regions}
\end{figure}

Having learned a highly predictive model, we would like to locate where the selected $k$-mers occur in the viral proteomes. We visualize the $k$-mer subsequences selected in a specific ADT by indicating elements of the mismatch neighborhood of each selected subsequence on the virus protein sequences. In Figure \ref{fig:predictive_regions}, the virus proteomes are grouped vertically by their label with their lengths scaled to $[0,1]$. Quite surprisingly, the predictive $k$-mers occur in regions that are strongly conserved among viruses sharing a specific host. Note that the representation we used for viral sequences retained no information regarding the location of each $k$-mer on the virus protein. Furthermore, these selected $k$-mers are significant as they are robustly selected by Adaboost for different choices of train / test split of the data, as shown in Figure \ref{fig:robust_regions}.

\section{Discussion}

We have presented a supervised learning algorithm that learns a model to classify viruses according to their host and identifies a set of highly discriminative oligopeptide motifs. As expected, the $k$-mers selected in the ADT for \emph{Picornaviridae} (Figure \ref{fig:predictive_regions}, \ref{fig:robust_regions}) and \emph{Rhabdoviridae} (Figure \ref{fig:predictive_regions_rhabdo}, \ref{fig:robust_regions_rhabdo}) are mostly selected in areas corresponding to the replicase motifs of the polymerase -- one of the most conserved parts of the viral genome. Thus, given that partial genomic sequence is normally the only information available, we could achieve quicker bioinformatic characterization by focusing on the selection and amplification of these highly predictive regions of the genome, instead of full genomic characterization and contiguing. Moreover, in contrast with generic approaches currently under use, such a targeted amplification approach might also speed up the process of sample preparation and improve the sensitivity for viral discovery. 

Over representation of highly similar viruses within the data used for learning is an important source of overfitting that should be kept in mind when using this technique. Specifically, if the data largely consist of nearly-similar viral sequences (e.g. different sequence reads from the same virus), the learned ADT model would overfit to insignificant variations within the data (even if 10-fold cross validation were employed), making generalization to new subfamilies of viruses extremely poor. To check for this, we hold out viruses corresponding to a particular subfamily (see Table S1 for subfamily annotation of the viruses used), run 10-fold cross validation on the remaining data and compute the expected fraction of misclassified viruses in the held-out subfamily, averaged over the learned ADT models. For \emph{Picornaviridae}, viruses belonging to the subfamilies \emph{Parechovirus} (0.47), \emph{Tremovirus} (0.8), \emph{Sequivirus} (0.5), and \emph{Cripavirus} (1.0) were poorly classified with misclassification rates indicated in parentheses. Note that the \emph{Picornaviridae} data used consist mostly of Cripaviruses; thus, the high misclassification rate could also be attributed to a significantly lower sample size used in learning. For \emph{Rhabdoviridae}, viruses belonging to \emph{Novirhabdovirus} (0.75) and \emph{Cytorhabdovirus} (0.77) were poorly classified. The poorly classified subfamilies, however, contain a very small number of viruses, showing that the method is strongly generalizable on average.

Other applications for this technique include identification of novel pathogens using genomic data, analysis of the most informative fingerprints that determine host specificity, and classification of metagenomic data using genomic information. For example, an alternative application of our approach would be the automatic discovery of multi-locus barcoding genes. Multi-locus barcoding is the use of a set of genes which are discriminative between species, in order to identify known specimens and to flag possible new species \cite{Seberg2009}. While we have focused on virus host in this work, ADTs could be applied straightforwardly to the barcoding problem, replacing the host label with a species label. Additional constraints on the loss function would have to be introduced to capture the desire for suitable flanking sites of each selected $k$-mer in order to develop the universal PCR primers important for a wide application of the discovered barcode \cite{Kress2008}.

\section*{Acknowledgments}
The authors thank Vladimir Trifonov and Joseph Chan for interesting suggestions and discussions.

\bibliographystyle{unsrt}
\bibliography{references}

\begin{thebibliography}{10}

\bibitem{Briese2009}
Thomas Briese, Janusz~T Paweska, Laura~K McMullan, Stephen~K Hutchison, Craig
  Street, Gustavo Palacios, Marina~L Khristova, Jacqueline Weyer, Robert
  Swanepoel, Michael Egholm, Stuart~T Nichol, and W~Ian Lipkin.
\newblock {Genetic detection and characterization of Lujo virus, a new
  hemorrhagic fever-associated arenavirus from southern Africa.}
\newblock {\em PLoS pathogens}, 5(5):e1000455, May 2009.

\bibitem{Williamson2008}
Shannon~J Williamson, Douglas~B Rusch, Shibu Yooseph, Aaron~L Halpern, Karla~B
  Heidelberg, John~I Glass, Cynthia Andrews-Pfannkoch, Douglas Fadrosh,
  Christopher~S Miller, Granger Sutton, Marvin Frazier, and J~Craig Venter.
\newblock {The Sorcerer II Global Ocean Sampling Expedition: metagenomic
  characterization of viruses within aquatic microbial samples.}
\newblock {\em PloS one}, 3(1):e1456, January 2008.

\bibitem{Touchon2008}
Marie Touchon and Eduardo P~C Rocha.
\newblock {From GC skews to wavelets: a gentle guide to the analysis of
  compositional asymmetries in genomic data.}
\newblock {\em Biochimie}, 90(4):648--59, 2008.

\bibitem{Duffy2008}
Siobain Duffy, Laura~A Shackelton, and Edward~C Holmes.
\newblock {Rates of evolutionary change in viruses: patterns and determinants.}
\newblock {\em Nature Reviews Genetics}, 9(4):267--276, 2008.

\bibitem{Pybus2009}
Oliver~G Pybus and Andrew Rambaut.
\newblock {Evolutionary analysis of the dynamics of viral infectious disease.}
\newblock {\em Nature reviews. Genetics}, 10(8):540--50, August 2009.

\bibitem{Leslie2004}
C~S Leslie, E~Eskin, A~Cohen, J~Weston, and W~S Noble.
\newblock {Mismatch string kernels for discriminative protein classification.}
\newblock {\em Bioinformatics}, 20(4):467--476, 2004.

\bibitem{Hamming1950}
R~W Hamming.
\newblock {Error detecting and error correcting codes}.
\newblock {\em Bell System Technical Journal}, 29:147----160, 1950.

\bibitem{Schapire1999}
R~E Schapire and Y~Singer.
\newblock {Improved boosting algorithms using confidence-rated predictions}.
\newblock {\em Machine Learning}, 37(3):297--336, 1999.

\bibitem{Busa2009}
R~Busa-Fekete and B~Kegl.
\newblock {Accelerating AdaBoost using UCB}.
\newblock In {\em JMLR: Workshop and Conference Proceedings, KDD cup 2009},
  volume~7, pages 111--122, 2009.

\bibitem{Freund1999}
Y~Freund and L~Mason.
\newblock {The Alternating Decision Tree Algorithm.}
\newblock In {\em Proceedings of the 16th International Conference on Machine
  Learning}, pages 124--133, 1999.

\bibitem{Altschul1990}
Stephen~F Altschul, Warren Gish, Webb Miller, Eugene~W Myers, and David~J
  Lipman.
\newblock {Basic Local Alignment Search Tool}.
\newblock {\em Journal of Molecular Biology}, 215:403--410, 1990.

\bibitem{Seberg2009}
Ole Seberg and Gitte Petersen.
\newblock {How many loci does it take to DNA barcode a crocus?}
\newblock {\em PloS one}, 4(2):e4598, January 2009.

\bibitem{Kress2008}
W~John Kress and David~L Erickson.
\newblock {DNA barcodes: genes, genomics, and bioinformatics.}
\newblock {\em Proceedings of the National Academy of Sciences of the United
  States of America}, 105(8):2761--2762, 2008.

\end{thebibliography}
\renewcommand{\thefigure}{S.\arabic{figure}}
\renewcommand{\thetable}{S.\arabic{table}}
\setcounter{figure}{0}  
\setcounter{table}{0}  
\section*{Supplementary Figures}
\begin{figure}[!ht]
    \begin{center}
        \includegraphics{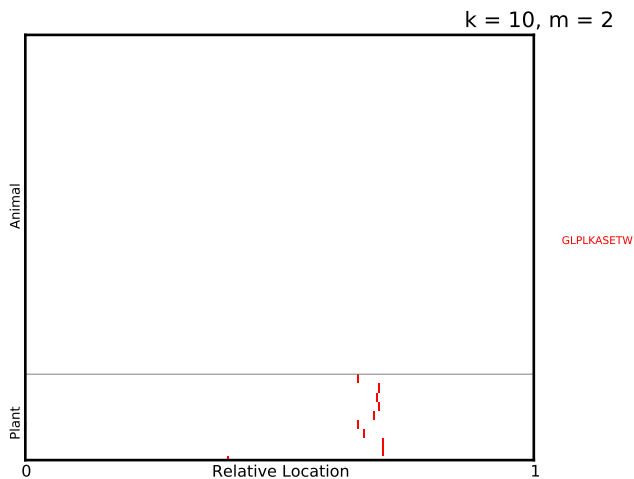}
    \end{center}
    \caption{{\bf Visualizing predictive subsequences for \emph{Rhabdoviridae}.} A visualization of the mismatch neighborhood of the $k$-mer selected in an ADT for \emph{Rhabdoviridae}, where $k = 10, m = 2$. The virus proteomes are grouped vertically by their label with their lengths scaled to $[0,1]$. Regions containing elements of the mismatch neighborhood of each $k$-mer are then indicated on the virus proteome.}
    \label{fig:predictive_regions_rhabdo}
\end{figure}
\begin{figure}[!ht]
    \begin{center}
        \includegraphics{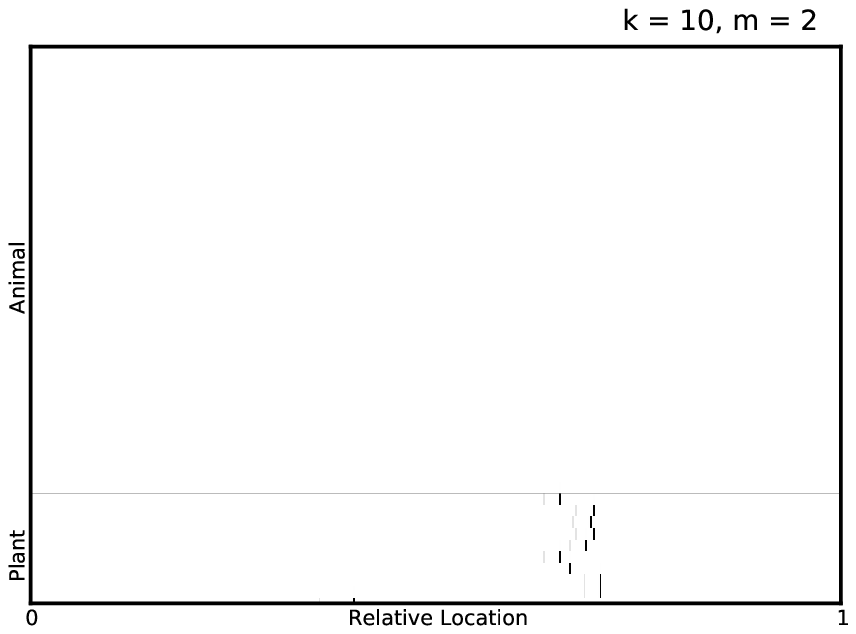}
    \end{center}
    \caption{{\bf Visualizing predictive regions for \emph{Rhabdoviridae}.} A visualization of the mismatch neighborhood of the $k$-mers selected in an ADT for \emph{Rhabdoviridae}, where $k = 10, m = 2$. The virus proteomes are grouped vertically by their label with their lengths scaled to $[0,1]$. Regions containing elements of the mismatch neighborhood of each $k$-mer are then indicated on the virus proteome.}
    \label{fig:robust_regions_rhabdo}
\end{figure}

\end{document}